\documentstyle[12pt]{article}

\newcommand{\beq}[1]{\begin{equation}\label{#1}}
\newcommand{\eeq}{\end{equation}}
\newcommand{\bear}[1]{\begin{eqnarray}\label{#1}}
\newcommand{\ear}{\end{eqnarray}}

\newcommand{\rf}[1]{(\ref{#1})}
\newcommand{\nl}{ {\hfill \break} }
\newcommand{\np}{ {\newpage } }

\newcommand{\Iff}{ {\Leftrightarrow } }

\newcommand{\imp}{\ {\Rightarrow }\ }

\newcommand{\R}{ \mbox{\rm I$\!$R} }
\newcommand{\Diff}{ \mbox{\rm Diff} }
\newcommand{\Out}{ \mbox{\rm Out} }
\newcommand{\Inn}{ \mbox{\rm Inn} }
\newcommand{\Aut}{ \mbox{\rm Aut} }
\newcommand{\obs}{ \mbox{\rm obs} }

\newcommand{\tr}{ \mbox{\rm tr} }

\begin{document} 
\vspace*{-1.5cm}
\centerline{\mbox{\hspace*{10cm}Prepr. Math. Inst.}}
\centerline{\mbox{\hspace*{10cm}Univ. Potsdam 96/17}}
\vspace*{0.3cm}
\begin{center} 
{\Large
 {\bf Regularized Algebraic Nets for 
\\
General Covariant QFT on Differentiable Manifolds\footnote
{\bf\em This work was financially supported by a 
DAAD fellowship.}
\\}}  
\vskip 0.35cm 

{\large {\bf 
Martin Rainer\footnote { e-mail:
mrainer@aip.de}
}}\\
\vskip 0.31 cm  

{Institut f\"ur Mathematik} \\
{ Universit\"at Potsdam, PF 601553} \\
{D-14415 Potsdam, Germany}  \\
{\vskip 0.13 cm}

{Institute for Studies in Physics and Mathematics} \\
P.O.Box 19395-5531, Tehran, Iran  \\
{\vskip 0.26 cm}

{\em June 1996}
\end{center}

\begin{abstract}
Quantum general relativity may be considered as generally covariant
QFT on differentiable manifolds, without any a priori metric structure.
The kinematically covariance group
acts by general diffeomorphisms  on the manifold and by
automorphisms on the isotonic net of $*$-algebras encoding the QFT,
while the algebra of observables is covariant under the dynamical subgroup 
of the general diffeomorphism group. 

Here, I focus on an algebraic implementation of
the dynamical subgroup of dilations. Introducing
an small and large scale cutoffs algebraically,  
their usual a priori conflict with general covariance
is avoided. Thereby, a commutant duality
between the minimal and maximal algebra
is proposed. This allows to extract the modular structure, which
is again related to the dilations.  
\end{abstract}
\vspace{0.21cm}
\np
\section{Introduction} 
{Observation procedures} 
represent the abstract kinematical framework
for {possible} preparations of measurements, while the {observables}
encode the kinds of questions one can ask from the physical system.
The importance of this distinction was observed by Ekstein \cite{Ek}
about 30 years ago. Since then practically any reasonable attempt
for a constructive quantum theory in general, and for a constructive
quantum field theory (QFT) in particular, 
is striving  for a consistent implementation
of observation procedures and observables. 

The covariance group of the observation procedures reflects the 
general (a priori) redundancy of their mathematical implementation. 
The more sophisticated the structure of the
observation procedures, the smaller the covariance group will be in general.
E.g. in \cite{MR} the kinematical observation procedures are given
by a network of discrete vertices of a specific Riemannian surface
embedded in a $3+1$-dimensional space-time $M$, whence the covariance
group is only that subgroup of $\Diff(M)$ which leaves this structure
invariant. In general, it is a difficult question, how much structure
might be put on the observation procedures. For the following however,
I will just follow a common philosophy of general covariance,
to impose as little a priori structure as possible.

In a concrete observation the kinematical covariance will be broken.
So in \cite{MR} a concrete local observation requires the 
explicit selection of one of many a priori equivalent vertices,
whence it breaks the covariance which holds for the network of vertices
as a whole.
In the examples of \cite{Ek} the kinematical covariance was assumed to be 
broken in a concrete observation by a dynamical interaction with external 
fields. 
However, irrespectively of the loss of 
covariance in a concrete observation, the action
of the covariance group may still be well defined on the observation 
procedures.
In any case, the loss of covariance in a concrete observation is related
to a specific structure of the state of the physical system.
\np
Let us examine now the consequences of this breaking of general covariance 
within an algebraic approach  
to a generally covariant quantum field theory. 
The first step in this direction was actually already done in \cite{FrHa},
and picked up further in \cite{Sal}.
The principle of locality
is at the heart of the constructive 
approach to quantum field theory \cite{HaKa}.
Here it is kept in form of the demand that, observation procedures 
correspond to {possible} preparations of 
localized measurements in finite regions. 
Note that finiteness is 
a purely topological notion.
I do not assume here  any priori notion of 
neither a metric, time, nor even a causal structure. 
Hence, on different regions there will be no a priori 
causal relations between observables. 
It was shown in \cite{Ba1} that, for
the net of subalgebras of a Weyl algebra, it is indeed possible to
work with a flexible notion of causality rather than with a rigidly given 
one.

Although  in principle it might be possible to construct a net together with
its underlying manifold from a partial order via inclusion
of the algebras themselves (cf. \cite{Ba2}),
we will start in Sect. 2 with a net of $*$-algebras on a 
differentiable manifold.
On this net, a physical state induces dynamical relations,
whence the algebra of observables is covariant under the dynamical subgroup 
of the general diffeomorphism group. 
The present examinations emphasize on
the dynamical subgroup of dilations.
Sect. 3 is devoted to the implementation of a
a small and large scale regularization indirectly, 
thus avoiding the usual direct conflict between 
cutoffs and general covariance.
A new commutant duality between the corresponding
minimal and maximal algebra is introduced.
In Sect. 4 this duality, together with the isotony property,
is used to extract the modular structure. The latter is 
related to local dilations on the net.
Sect. 5 concludes with a brief discussion of some possible
implications of the proposed structure for quantum general relativity
and a posteriori notions of time and causality.
\section{Generally covariant nets of algebras} 
\setcounter{equation}{0}
Let us consider a net on a differentiable manifold $M$,
which associates to each open set 
${\cal O} \in M$ a $*$-algebra ${\cal A}({\cal O})$
such that isotony, 
\beq{iso}
{\cal O}_1\subset {\cal O}_2\imp
{\cal A}({\cal O}_1)\subset {\cal A}({\cal O}_2),
\eeq
holds.
The following investigations might be seen as an attempt to understand
some aspects of quantum field theory (QFT) on differentiable manifolds. 
This is indeed a very promising approach 
to quantum general relativity \cite{Ro}. 
Selfadjoint elements of ${\cal A}({\cal O})$ 
may be interpreted as {\em observation procedures},
i.e. possible prescriptions for laboratory measurements in $\cal O$. 

There should not be any a priori relations between observation procedures 
associated with disjoint regions. In other words, the net
${\cal A}:={\bigcup_{\cal O}} {\cal A}({\cal O})$ has to be
free from any relations which exceed its mere definition.\\

This interpretation allows us to extend the $\Diff(M)$ covariance  
{}from the underlying manifold $M$ to the net of algebras,
on which $\Diff(M)$ then acts by automorphisms, 
i.e. each diffeomorphism
$\chi\in \Diff(M)$ induces an automorphism 
$\alpha_{\chi}$ of the observation procedures such that
\begin{equation}
\alpha_{\chi}({\cal A}({\cal O}))={\cal A}(\chi({\cal O})).
\label{I1}  
\end{equation}
The state of a physical system is mathematical described by a 
positive linear functional  ${\omega}$ on $\cal A$. 
Given the state ${\omega}$, one
gets via the GNS construction a representation $\pi^{\omega}$ of $\cal A$ 
by a net of operator algebras on a Hilbert space ${\cal H}^{\omega}$ with 
a cyclic vector $\Omega^{\omega}\in {\cal H}^{\omega}$. The
GNS representation $(\pi^{\omega}, {\cal H}^{\omega}, \Omega^{\omega})$ 
of any state $\omega$ has a so called folium ${\cal F}^{\omega}$, 
given as the family of those states $\omega_\rho:=\tr\rho\pi^{\omega}$ 
which are defined by positive trace class
operators $\rho$ on ${\cal H}^{\omega}$. 

Once a physical state $\omega$ has been specified, one can consider in each
algebra ${\cal A}({\cal O})$ the equivalence relation
\begin{equation}
A\sim B \ \ :\Iff \ \ 
{\omega}^{\prime}(A-B)=0,\ \ \forall  {\omega}^{\prime}\in{\cal F}^{\omega}.
\end{equation}
These equivalence relations generate a two-sided ideal 
${\cal I}^{\omega}({\cal O}):=\{A\in{\cal A}({\cal O})\vert
{\omega}^{\prime}(A)=0\}$ in ${\cal A}({\cal O})$.
The algebra of {\em  observables} 
${\cal A}^{\omega}_{\obs}({\cal O}):=\pi^{\omega}({\cal A}({\cal O}))$ may
be constructed from
the algebra of observation procedures ${\cal A}({\cal O})$ by taking 
the quotient
\begin{equation}
{\cal A}^{\omega}_{\obs}({\cal O}):=
{\cal A}({\cal O})/{\cal I}^{\omega}({\cal O}).
\end{equation}
Since any diffeomorphism $\chi\in \Diff(M)$ induces an 
automorphism $\alpha_{\chi}$ of the observation procedures,
one may ask whether, for a given state $\omega$,  
the action of $\alpha_{\chi}$ will leave the net 
${\cal A}^{\omega}_{\obs}:=
{\bigcup_{\cal O}} {\cal A}^{\omega}_{\obs}({\cal O})$
of observables invariant, with an action of the form
\begin{equation}
\alpha_{\chi}({\cal A}^{\omega}_{\obs}({\cal O}))=
{\cal A}^{\omega}_{\obs}(\chi({\cal O})).
\label{D1}
\end{equation}
In order for this to be possible, the ideal
${\cal I}^{\omega}({\cal O})$ must transform covariantly, i.e.
the diffeomorphism ${\chi}$ must satisfy
\begin{equation}
\alpha_{\chi}({\cal I}^{\omega}({\cal O}))=
{\cal I}^{\omega}(\chi({\cal O})).
\label{D2}
\end{equation}
Hence, the algebra of 
observables, constructed with respect to
the folium ${\cal F}^\omega$, does no longer exhibit the 
kinematical  $\Diff(M)$ symmetry of the observation procedures. 
The symmetry of the observables is dependent on (folium of) the state
$\omega$. 
Therefore, the selection of a folium of states
${\cal F}^\omega$, induced by the actual choice of a state $\omega$,
results immediately in a breaking of the $\Diff(M)$ symmetry.
The resulting effective symmetry group, also briefly called  
the {\em dynamical group} of the state $\omega$,
is given by the subgroup of those diffeomorphisms which satisfy 
the constraint condition (\ref{D2}).  
An automorphisms $\alpha_{\chi}$ is called {\em dynamical}
(w.r.t. the given state $\omega$) if it satisfies (\ref{D2}).

The remaining dynamical symmetry group, depending
on the folium ${\cal F}^\omega$ of states related to $\omega$,
has two main aspects which we have to examine
if we actually want to specify the physically admissible states:
Firstly, it is necessary to specify its state dependent automorphic 
algebraic action on the net of observables. Secondly, one has to
find a geometric interpretation for the group and its action on $M$.

If we consider the dynamical group as an {\em inertial}, 
and therefore global, manifestation of dynamically ascertainable 
properties of observables, 
then its (local) action should be correlated with (global) 
operations on the whole net of observables.
This implies that at least some of the dynamical
automorphisms $\alpha_\chi$ are not inner.
(For the case of causal nets of algebras 
it was actually already shown in \cite{Wo} 
that, under some additional assumptions,
the automorphisms of the algebras are in general not inner.)

Note that one might consider instead of the net of observables
${\cal A}^{\omega}_{\obs}({\cal O})$  
the net of associated von Neumann algebras
${\cal R}^{\omega}_{\obs}({\cal O})$, which can be defined even for
unbounded ${\cal A}^{\omega}_{\obs}({\cal O})$, if we take 
{}from the modulus of the von Neumann closure  
$({{\cal A}^{\omega}_{\obs}}({\cal O}))''$ 
all its spectral projections \cite{FrHa}.
Then the isotony (\ref{iso}) induces a likewise isotony of the net 
${\cal R}^{\omega}_{\obs}:=
{\bigcup_{\cal O}} {\cal R}^{\omega}_{\obs}({\cal O})$
of von Neumann algebras.

\section{Algebraic small and large scale regularization} 
\setcounter{equation}{0}
In the following I want to exhibit a possibility to introduce
both, small and large scale cutoff regularizations
on the net of von Neumann algebras.
This essentially exploits a local partial ordering on the net,
which is induced by the isotony property.

Let us now make use of the given ($C^\infty$) topological structure of
$M$ and choose at point $x\in M$ a topological basis
of nonzero open sets 
${\cal O}^{x}_s\ni x$,
parametrized by a real parameter $s$ with $0<s<\infty$, 
such that
\begin{equation}
s_1<s_2 \quad \Iff \quad {\cal O}^{x}_{s_1}{\subset} {\cal O}^{x}_{s_2} 
\label{inc}
\end{equation}
and 
\begin{equation}
s\to 0 \quad \Iff \quad {\cal O}^{x}_{s}{\to} \emptyset .
\end{equation}
Let us now further restrict the parameter $s$ such that
$0<s_{\min,x}<s<s_{\max,x}<\infty$ and 
assume 
\bear{univers}
s_{\min,x}=s_{\min} ,\quad s_{\max,x}=s_{\max} \qquad \forall x\in M .   
\ear
Then,
for each $x\in M$, open sets ${\cal O}^{x}_{s}$ with 
$s\in ]s_{\min},s_{\max}[$ generate local cobordisms between
$\partial {\cal O}^{x}_{s_{\min}}$
and $\partial {\cal O}^{x}_{s_{\max}}$, and
the isotony property 
implies that 
\begin{equation}
{\cal R}^{\omega}_{\obs}({\cal O}^{x}_{s_{\min}})
{\ \subset\ } {\cal R}^{\omega}_{\obs}({\cal O}^{x}_{s})
{\ \subset\ } {\cal R}^{\omega}_{\obs}({\cal O}^{x}_{s_{\max}}).
\end{equation}
The key step is now to impose a commutant duality relation between
the inductive limits given by the minimal and maximal algebras,
\begin{equation}
{\cal R}^{\omega}_{\obs}({\cal O}^{x}_{s_{\min}})
=\left({{\cal R}^{\omega}_{\obs}({\cal O}^{x}_{s_{\max}})}\right)',
\label{dmin}
\end{equation}
where ${\cal R}'$ denotes the commutant of ${\cal R}$ within some
${\cal R}_{\max}\supset {\cal R}$. 
Then the bicommutant theorem (${\cal R}''={\cal R}$) 
implies that likewise also
\begin{equation}
{\cal R}^{\omega}_{\obs}({\cal O}^{x}_{s_{\max}})
=\left({{\cal R}^{\omega}_{\obs}({\cal O}^{x}_{s_{\min}})}\right)'.
\label{dmax}
\end{equation}
If one now demands that all maximal (or all minimal) algebras
are isomorphic to each other, independently of the choice of $x$ and the
open set ${\cal O}^{x}_{s_{\max}}$ 
(resp. ${\cal O}^{x}_{s_{\min}}$),
then by (\ref{dmin}) (resp. (\ref{dmax})) also all minimal (resp. maximal)
algebras are isomorphic to each other.  
I then denote the universal minimal resp. maximal algebra as 
${\cal R}^{\omega}_{{\min}}$ and ${\cal R}^{\omega}_{{\max}}$ 
respectively. 
In the following the commutant will always been 
taken within ${\cal R}^{\omega}_{{\max}}$.
Then, the duality (\ref{dmin}) implies that 
${\cal R}^{\omega}_{{\min}}$ is Abelian.

By isotony and (\ref{inc}) together with \rf{univers}, 
the mere existence of 
${\cal R}^{\omega}_{\min}$ 
resp. ${\cal R}^{\omega}_{\max}$
fixes  already a common size (as measured by the parameter $s$) of
all sets ${\cal O}^{x}_{s_{\min}}$ resp. ${\cal O}^{x}_{s_{\max}}$
independently of $x\in M$.    
So in this case $s_{\min}$ and $s_{\max}$ really denote 
an universal small resp. large scale cutoff.
Note that, in the context of Sect. 2, the universality assumption 
\rf{univers} is indeed nontrivial, because local diffeomorphisms 
consistent 
with the structure above must preserve $s_{\min}$,  $s_{\max}$,
and the monotony of the ordered set $]s_{\min},s_{\max}[$.
The number $s\in]s_{\min},s_{\max}[$
parametrizes the partial order of the net of algebras
spanned between the inductive limits 
${\cal R}^{\omega}_{\min}$ and ${\cal R}^{\omega}_{\max}$.

Although in local QFT usually the supports of an algebra and
its commutant are not at all related,
it might be nevertheless instructive to consider the
case where the algebras satisfy 
\begin{equation}
\label{cominc}
\left({{\cal R}^{\omega}_{\obs}({\cal O}_s^{x})}\right)'
\subset {\cal R}^{\omega}_{\obs}({\cal O}_s^{x}) .
\end{equation}
Then, with the center of ${\cal R}^{\omega}_{\obs}({\cal O}_s^{x})$
defined as 
${\cal Z}\left({\cal R}^{\omega}_{\obs}({\cal O}_s^{x})\right):=
{\cal R}^{\omega}_{\obs}({\cal O}_s^{x})\cap
\left({{\cal R}^{\omega}_{\obs}({\cal O}_s^{x})}\right)'$,
one obtains 
${\cal Z}\left({\cal R}^{\omega}_{\obs}({\cal O}_s^{x})\right)=
\left({{\cal R}^{\omega}_{\obs}({\cal O}_s^{x})}\right)'=
{\cal Z}\left(({{\cal R}^{\omega}_{\obs}({\cal O}_s^{x})})'\right)$,
and especially
${\cal Z}\left({\cal R}^{\omega}_{\max}\right)=
{{\cal R}^{\omega}_{\min}}=
{\cal Z}\left({\cal R}^{\omega}_{\min}\right)$.
So, for a pair of commutant dual algebras satisfying
Eq. (\ref{cominc}), the smaller one is always
Abelian, namely it is the center of the bigger one.
With (\ref{cominc}), the isotony of the net implies the existence 
of an algebra ${\cal Z}^{\omega}$ which is {\em maximal Abelian}, 
in other words
commutant selfdual, satisfying ${\cal Z}^{\omega}=({\cal Z}^{\omega})'=
{\cal Z}({\cal Z}^{\omega})$. This algebra is given explicitly via the
Abelian net of all centers, ${\cal Z}^{\omega}
:={\bigcup_{\cal O}}{\cal Z}\left({\cal R}^{\omega}_{\obs}({\cal O})\right)$.
${\cal Z}^{\omega}$, located on an underlying set ${\cal O}^{x}_{s_{z}}$
of intermediate size s.th. $s_{\min}<s_z<s_{\max}$,  separates
the small Abelian algebras 
${{\cal R}^{\omega}_{\obs}({\cal O}_s^{x})}=
{\cal Z}\left({\cal R}^{\omega}_{\obs}({\cal O}_s^{x})\right)$,
with $s\leq s_z$, from larger
non-Abelian algebras 
${{\cal R}^{\omega}_{\obs}({\cal O}_s^{x})}=
\left({\cal Z}({\cal R}^{\omega}_{\obs}({\cal O}_s^{x}))\right)'$,
with $s>s_z$. 

For a net subject to (\ref{cominc}), its lower end is Abelian, 
whence observations on small regions  with $s\leq s_z$
are expected to be rather classical.
Nevertheless, for increasing size  $s>s_z$, 
there might well exist a non-trivial
quantum (field) theory 
(in \cite{Wo} it was shown that, for  causal nets, the
algebras of QFT are not Abelian and not finite-dimensional).
For quantum general relativity there might indeed be a kinetic substructure
\cite{Sal1}. Classical elementary constituents of the latter naturally
span an Abelian algebra.
It is interesting in this context that the Abelian part of the 
loop algebra of quantum general relativity
provides indeed the classical spectrum \cite{Ro,AshIsh}.
\section{Modular structure and dilations} 
\setcounter{equation}{0}
If we consider the small and large scale cutoffs as introduced above,
it should be clear that only those regions (\ref{dc})
of size $s\in [s_{\min},s_{\max}]$ are admissible for measurement.  
The commutant duality between 
${\cal R}^{\omega}_{\min}$ and ${\cal R}^{\omega}_{\max}$
inevitably yields large scale correlations 
in the structure of any physical state $\omega$ on
any admissible region ${\cal O}^{x}_{s}$ of measurement at $x$.
Let us assume here that $\omega$ is properly correlated,
i.e. the GNS vector $\Omega^\omega$ is already cyclic under 
${\cal R}^{\omega}_{\min}$. 
Then, by duality,  it is separating for 
${\cal R}^{\omega}_{\max}={{\cal R}^{\omega}_{\min}}'$.
Furthermore $\Omega^\omega$  is also cyclic under  
${\cal R}^{\omega}_{\max}$, and hence
separating for ${\cal R}^{\omega}_{\min}$.

So $\Omega^{\omega}$ is a cyclic and separating vector for 
${\cal R}^{\omega}_{\min}$ and ${\cal R}^{\omega}_{\max}$, 
and by isotony also for any
local von Neumann algebra ${\cal R}^{\omega}_{\obs}({\cal O}^{x}_{s})$.

As a further consequence, on any region ${\cal O}^{x}_{s}$, 
the Tomita operator $S$ and and its conjugate $F$ 
can be defined densely by
\begin{equation}
S A \Omega^{\omega}:= A^{*} \Omega^{\omega} \ \ \mbox{for}\ \ A\in 
{\cal R}^{\omega}_{\obs}({\cal O}^{x}_{s})
\label{T0}
\end{equation}
\begin{equation}
F B \Omega^{\omega}:=B^{*} \Omega^{\omega}
 \ \ \mbox{for}\ \ B\in 
{{\cal R}^{\omega}_{\obs}({\cal O}^{x}_{s})}' . 
\label{T1}
\end{equation}
The  closed Tomita operator $S$ has a polar decomposition 
\begin{equation}
S=J\Delta^{1/2},
\label{T2}\end{equation}
where  $J$ is antiunitary and $\Delta:=FS$ is the self-adjoint, positive 
modular operator.
The Tomita-Takesaki theorem \cite{Ha} provides us with a one-parameter 
group of state dependent automorphisms $\alpha^{\omega}_t$ on 
${\cal R}^{\omega}_{\obs}({\cal O}^{x}_{s})$,
defined by
\begin{equation}
\alpha^{\omega}_t (A)= \Delta^{-it}\ A\ \Delta^{it},   \ \ \mbox{for}\ \ 
A\in{\cal R}^{\omega}_{\max}.
\label{T3}
\end{equation}
So, as a consequence of commutant duality and isotony assumed above,  
we obtain here a strongly continuous unitary 
implementation  of the modular group of $\omega$,
which is defined by  the $1$-parameter family of automorphisms (\ref{T3}),
given as conjugate action of operators
$e^{-it\ln\Delta}$, ${t\in\R}$.
By (\ref{T3}) the modular group, for a state $\omega$  
on the net of von Neumann algebras, 
defined by ${\cal R}^{\omega}_{\max}$,  
might be considered as a $1$-parameter subgroup of the dynamical group.
Note that, with Eq. (\ref{T1}), in general,
the modular operator $\Delta$ is not located on
${\cal O}^{x}_{s}$. Therefore,  in general, the modular automorphisms 
(\ref{T3}) are not inner.
It is known (see e.g. \cite{BaWo}) that the modular automorphisms act as 
{inner} automorphisms, iff the von Neumann algebra 
${\cal R}^{\omega}_{\obs}({\cal O}^{x}_{s})$
generated by $\omega$
contains only semifinite factors, 
i.e. factors of type I and II. In this case  $\omega$ is a semifinite
trace.

Above we considered concrete von Neumann algebras
${\cal R}^{\omega}_{\obs}({\cal O}^{x}_{s})$, which are in fact 
operator representations of an abstract von Neumann algebra
${\cal R}$ on a GNS Hilbert space ${\cal H}^{\omega}$ w.r.t.
a faithful normal state $\omega$.
In general, different faithful normal states generate different 
concrete von Neumann algebras and different modular automorphism groups
of the same abstract  von Neumann algebra.

The outer modular automorphisms form the cohomology group  
$\Out {\cal R}:=\Aut {\cal R}/\Inn {\cal R}$
of modular automorphisms modulo inner modular automorphisms. 
This group is characteristic
for the types of factors contained in von Neumann algebra $\cal R$
(cf. \cite{Co}).
Per definition $\Out {\cal R}$ is trivial for inner automorphisms.
Factors of type III${}_1$ yield $\Out {\cal R}=\R$.

In the case of thermal equilibrium states, 
corresponding to factors 
of type III${}_1$ (see \cite{Ha}), there is a distinguished
$1$-parameter group of outer modular automorphisms,
which is a subgroup of the dynamical group. 

Looking for a geometric interpretation for this subgroup, 
parametrized by $\R$, it should not be a coincidence
that our partial order defined above is parametrized
by open intervals
(namely $]s_{\min},s_{\max}[$ for the full net
and, in the case of \rf{cominc}, $]s_{z},s_{\max}[$ 
for the non-Abelian part), 
and hence diffeomorphically likewise by $\R$.  
This way, dilations  of the open sets ${\cal O}^{x}_{s}$
within the open interval
may give a geometrical meaning to the $1$-parameter group of outer modular 
automorphisms of thermal equilibrium states. 
Even more, this might provide a perspective
for understanding the thermal time hypothesis of \cite{CoRo}.   
Indeed, a local equilibrium state might be characterized
as a KMS state  (see \cite{Ha,BrRo}) over the algebra of observables 
on a (suitably defined) double cone, whence the $1$-parameter modular group 
in the KMS condition might be related to the time evolution.
Note that, for double cones, a partial order
can be related to the split property of the algebras (see also \cite{Ba2}).

\section{Discussion} 
\setcounter{equation}{0}
A geometric action of the modular group
might be obtained by relating the thermal time
to the geometric notion of dilations of the open sets. 
For any $x\in M$, the parameter $s$ measures the extension
of the sets ${\cal O}^{x}_{s}$. 
As accessability regions for a local 
measurement in $M$, these sets naturally increase with time. 
Hence it is natural to suggest (at least for $s>s_z$) 
that the parameter $s$ might be
related to a (thermal) time $t$
such that, for any set ${\cal O}^{x}_{s}$,
$s>s_{\min}$, we have $t<s$ within the set and 
$t=s$ on the boundary $\partial{\cal O}^{x}_{s}$. 

For the ultralocal case 
($s_{\min}\to 0$, without UV cutoff), 
in \cite{Keyl} 
a construction of the causal structure for a space-time from 
the corresponding net of operator algebras was given.
Let us consider here the (a priori given) underlying manifold $M$ of the net.
Locally around any point $x\in M$ one may induce  open 
double cones  as the 
pullback of the standard double cone, which in fact is 
the conformal model of  Minkowski space.
These open double cones then carry natural
notions of time and causality, 
which are preserved under dilations.  
Therefore it seems natural to introduce locally around any $x\in M$ 
a causal structure and time by specializing the open sets to be
open double cones ${\cal K}^{x}_{s}$ located at $x$, with
time-like extension $2s$ between the ultimate past event $p$ and
the ultimate future event $q$ involved in any measurement in 
${\cal K}^{x}_{s}$ at $x$ 
(time $s$ between $p$ and $x$, and likewise between $x$ and $q$). 
Since the open double cones form a basis for the local topology of $M$, 
we might indeed consider equivalently the net of algebras located
on open sets 
\beq{dc}
{\cal O}^{x}_{s}:={\cal K}^{x}_{s}.
\eeq
Although some (moderate form of) locality might be indeed an 
indispensable principle within any reasonable theory of observations,
it is nevertheless an important but difficult question, 
under which consistency conditions
a local notion of time and causality might be extended 
{}from nonzero local environments of individual points to global regions.
This is of course also related to the non-trivial open question,
how open neighbourhoods of different points $x_1\neq x_2$
should be related consistently. Although, in a radical attempt to avoid some
part of these difficulties, one might try to replace the notion of points
and local regions by more abstract algebraic concepts,  
a final answer to these questions has not yet been found.
At least it seems natural that,
on manifolds with no a priori
causal structure, the net should satisfy  
a disjoint compatibility condition,
\begin{equation}\label{discom}
{\cal O}_1\cap {\cal O}_2=\emptyset
\quad \imp \quad [{\cal A}({\cal O}_1), {\cal A}({\cal O}_2)]=0 .
\end{equation}
This condition is e.g. also 
satisfied for Borchers algebras. 
Of course the inverse of (\ref{discom})
is not true in general.

Moreover, it is not yet clear whether (\ref{cominc})
is not a too strange condition (although we should of course
be aware that the feature of quantum general relativity might indeed
be very different from that of usual QFT).
(\ref{cominc}) makes sense, if
further investigation are able to proof the
existence of an Abelian substructure under reasonable conditions.

Note however that only the commutant duality
(\ref{dmin}), but not  (\ref{cominc}), was essential for
the extraction of the modular group.
If we assume the presence of factors of type
III${}_1$ in our von Neumann algebras (in the case of (\ref{cominc})
for a size $s>s_z$), or likewise the
existence of local equilibrium states, the choices
for time and causality, made above on the basis of
the partial order and related dilations are apparently natural. 
\nl\nl
{\large\bf Acknowledgement}
\nl\nl
This work was elaborated during summer 1996 visiting the IPM in Tehran
with the generous support of a DAAD fellowship. 
I am grateful to H. Salehi
for plenty of valuable discussions in a pleasant atmosphere
during that time.
Helpful comments by H. Baumg\"artel and  K. Fredenhagen 
and others are gratefully acknowledged.
\np

\end{document}